\documentclass[letterpaper,english,aps,preprint,superscriptaddress]{revtex4-1}
\usepackage[T1]{fontenc}
\usepackage[latin9]{inputenc}
\usepackage{color}
\usepackage{babel}
\usepackage{array}
\usepackage{float}
\usepackage{multirow}
\usepackage{amsmath}
\usepackage{amssymb}
\usepackage{graphicx}
\usepackage[unicode=true,pdfusetitle,
 bookmarks=true,bookmarksnumbered=false,bookmarksopen=false,
 breaklinks=false,pdfborder={0 0 1},backref=false,colorlinks=true]
 {hyperref}
\usepackage{breakurl}
\usepackage{tikz}
\usetikzlibrary{patterns}
\makeatletter

\providecommand{\tabularnewline}{\\}

\makeatother

\begin{document}

\preprint{}

\title{A Search for R-Parity Violated Scalar Tau-neutrino ($\widetilde{\nu_{\tau}}$)
at the Future Circular Collider}

\author{Yusuf Oguzhan Günaydin}
\email{yusufgunaydin@gmail.com}
\affiliation{Kahramanmaras Sütcü Imam University, Department of Physics, 46100, Kahramanmaras, Turkey }
 
 \author{Sehban Kartal}
\email{sehban@istanbul.edu.tr}
\affiliation{Istanbul University, Department of Physics, 34000, Istanbul,Turkey}

\author{Yunus Emre Okyayli}
\email{yunusokyayli@gtu.edu.tr}
\affiliation{Gebze Technical University, Department of Physics, 41400, Gebze, Kocaeli}

\author{Carmine Elvezio Pagliarone}
\email{carmine.pagliarone@lngs.infn.it}
\affiliation{INFN \textendash{} Laboratori Nazionali del Gran Sasso, Assergi (L\textquoteright Aquila) I-67100, Italy}
\affiliation{Dipartimento di Ingegneria Civile e Meccanica, Università degli Studi di Cassino e del Lazio Meridi- onale, Cassino I-03043, Italy}

\author{Mehmet Sahin}
\email{mehmet.sahin@usak.edu.tr}
\affiliation{Usak University, Department of Physics, 64200, Usak, Turkey}

\author{Saleh Sultansoy}
\email{ssultansoy@etu.edu.tr}
\affiliation{TOBB University of Economics and Technology, Department of Material Science and Nanotechnology Engineering,06560, Ankara, Turkey }
\affiliation{ANAS Institute of Physics, Baku, Azerbaijan}

\date{\today}
\begin{abstract}
Future Circular Collider is one of the next generation energy-frontier
proton-proton colliders with 100 TeV center of mass energy and promising
very high luminosity. FCC will be an extraordinary machine for searching
Beyond the Standard Model physics because of its very high center
of mass energy and luminosity. Searching supersymmetric particles
via R-parity violated interactions could be listed in promising BSM
physics at the FCC. Scalar tau neutrino is one of the most interesting
predictions. A search for $\widetilde{\nu}_{\tau}$ neutrino decaying
into $e\mu$ final state via R-parity violated interactions has been
conducted at the FCC-pp. It is seen from this research that FCC-pp
will be able to discover $\widetilde{\nu}_{\tau}$ up to 28.8 TeV,
observe 32.0 TeV and exclude 34.5 TeV mass values by taking $\lambda_{311}^{\prime}=0.11$
and $\lambda_{312}=\lambda_{321}=0.07$ at $\mathcal{L}_{int}=17500\,fb^{-1}$.
FCC-pp, also, will allow to examine very low values of the Yukawa
coupling constants which cause R-parity violation interactions. It
is obviously seen that FCC-pp will give an opportunity for detailed
research on scalar tau neutrino production and decay via R-parity
violated interactions. 
\end{abstract}



\maketitle

\section{\label{sec:intro}Introduction}

The theory of supersymmetry (SUSY) provides general form of the space-time
symmetries in quantum field theory that allows transformation bosons
into fermions and vise versa \citep{HABER198575,kane2000supersymmetric,pdg2016}.
In other words, a new fermionic and bosonic partner for each elementary
Standard Model (SM) fermions and bosons exists as well as an additional
Higgs boson doublet. These new particles are called gluino, gaugino,
sleptons, squarks and higgsino. R-parity is a basic concept of baryon
(B) and lepton (L) number conservation in particle interactions. R-parity
quantum number is described as $R=(-1)^{3(B-L)+2S}$ \citep{pdg2016}
or equivalently $R=(-1)^{3B+L+2S}$ \citep{BARBIER20051} where S
is spin quantum number. R-parity quantum number can be $\pm1$ that
if $R=+1$, particles are called R-even particles and include the
SM particles, else $R=-1$ means R-odd particles and include supersymmetric
particles. The conservation of R-parity in collision or decay processes
has fundamental consequences on supersymmetric phenomenology: supersymmetric
particles have to be produced in pairs which generally unstable particles
decay into lightest states. At the end of the process, a heavy unstable
supersymmetric particle decays into the lightest supersymmetric particles
(LSP) which must be in stable state due to R-parity invariance. R-parity,
however, is not conserved in such a states called R-parity violation
(RPV) that allows to LSP decay into SM particles. RPV can lead to
important phenomenological potential at particle physics \citep{DIMOPOULOS1988210,barger1989,barger1995,Kalinowski1997,Grossman1999,BARSHALOM2001297,Ghosh2002,BARBIER20051,Sierra_2005,Cakir_2012,Sahin_2013,Berger2015,Faroughy2015,Allanach2016,Chakraborty2016}. 

Right handed sneutrinos are considered as LSP \citep{Alan_2004,BARBIER20051,Khachatryan2016}
so if RPV occurs, sneutrinos decay into ordinary SM particles. Consequently,
sneutrinos can be observed by searching final state SM particles.
Then, plenty of experimental collaboration have been doing research
on sneutrinos via considering RPV interactions \citep{Abbiendietal.2000,Acosta2003,Abulencia2006,Abazov2008,abazov2010,Aaltonen2010,JACKSON_2012,Aaboud2016,Sirunyan2018,Aaboud:2018}.
The most recent ATLAS experiment results showed that mass limits of
$\widetilde{\nu_{\tau}}$ for $e\mu$, $e\tau$, and $\mu\tau$ final
states in $pp$ collision are 3.4, 2.9, and 2.6 TeV, respectively.
Coupling constants were taken as $\lambda_{311}^{'}=0.11$ and $\lambda_{312}=\lambda_{321}=0.07$
\citep{Aaboud:2018}. Lastly, CMS collaboration introduced experimental
results that $\widetilde{\nu_{\tau}}$ mass are excluded below 1.7
and 3.8 TeV for $e\mu$ final state in $pp$ collision while coupling
constants were taken as $\lambda_{312}=\lambda_{321}=\lambda_{311}^{'}=0.01$
and $0.1$, respectively \citep{Sirunyan2018}. 

In this work, we researched resonant production of stauneutrino ($\widetilde{\nu_{\tau}}$)
in $pp$ collisions at Future Circular Collider (FCC) with $\sqrt{{s}}=100$
TeV for $e\mu$ final state via both producing and decaying with R-parity
violation (RPV) interactions. Following sections are basic parameters
of FCC in section \ref{sec:Future-Circular-Collider}, super potential,
decay width and cross section in section \ref{sec:Superpotential,-Decay-Width},
signal and background analysis in section \ref{sec:Signal---Background}
and conclusion in section \ref{sec:Conclusion}. 

\section{Future Circular Collider\label{sec:Future-Circular-Collider}}

\begin{figure}[h]
\includegraphics[scale=0.5]{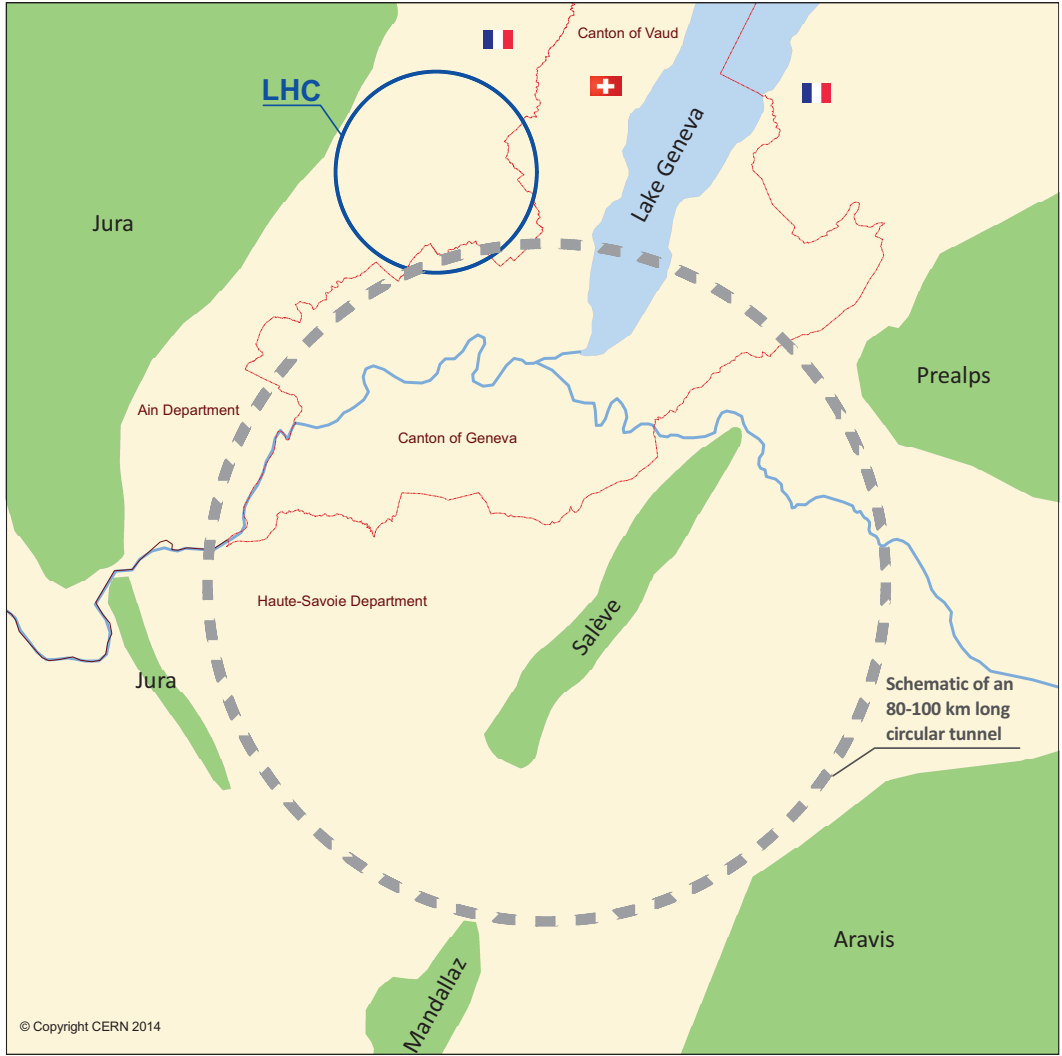}

\caption{Future Circular Collider's proposed layout.\label{fig:FCC}}

\end{figure}

Currently, the Large Hadron Collider (LHC) that has 13 TeV center
of mass energy provides proton-proton collisions at CERN. The Future
Circular Collider (FCC) is a post-LHC project that is proposed as
a new energy-frontier machine to be build at CERN (see Fig. \ref{fig:FCC}). 

\begin{table}[h]
\caption{FCC-pp main parameters for Phase I and Phase II.\label{tab:FCC-hh}}

\begin{tabular*}{16cm}{@{\extracolsep{\fill}}lcc}
\hline 
Parameters & Phase I  & Phase II\tabularnewline
\hline 
Circumferences (km) & \multirow{1}{*}{100} & 100\tabularnewline
Beam Energy (TeV) & 50 & 50\tabularnewline
Center of Mass Energy (TeV) & \multirow{1}{*}{100} & 100\tabularnewline
Peak Luminosity {[}$10^{34}\,cm^{-2}s^{-1}${]} & 5.1 & 29\tabularnewline
Integrated Luminosity {[}$fb^{-1}/yr${]} & $\geq$250 & $\geq$1000\tabularnewline
Operation Time {[}years{]} & 10 & 15\tabularnewline
Total Integrated Luminosity {[}$fb^{-1}${]} & 2500 & 17500\tabularnewline
\hline 
\end{tabular*}
\end{table}

FCC is designed with three options \citep{benedikt2016towards}: (a)
a 100-TeV proton-proton collider (FCC-pp) \citep{ball2014,benedikt2015fcc},
(b) proton-electron collider (FCC-he) \citep{benedikt2015future}
and (c) electron-positron collider (FCC-ee) \citep{wenninger2014future}.
Center of mass energy will be 100 TeV for FCC-pp but the other two
options have different scenarios such as FCC-ee might have center
of mass energy between 61 and 350 GeV and FCC-he's center of mass
energy varies from 3.46 to 31.6 TeV \citep{Abelleira_Fernandez_2012,benedikt2016towards,ACAR201747}.
Main collider tunnel is planned about 80 or 100 km circumferences
near Geneva that allows to reach very high integrated luminosity per
year in two phases. Integrated luminosity will be 250 $fb^{-1}$ per
year during the Phase I and 1000 $fb^{-1}$ per year during the Phase
II. FCC main parameters are summarized in Tab. \ref{tab:FCC-hh} that
total integrated luminosity depict cumulative values of the operation
time for two phases: 2500 $fb^{-1}$ (Phase I) and 17500 $fb^{-1}$
(Phase I + Phase II). It seems that FCC will be a very high luminosity
frontier machine that is going to play important role for new physics
researches. 

\section{Superpotential, Decay Width and Cross Section\label{sec:Superpotential,-Decay-Width}}

Superpotential of RPV interactions for $\tau$ sneutrino is defined
by Eq. \ref{eq:superpotential} \citep{BARBIER20051}, 

\begin{equation}
W_{RPV}=\frac{1}{2}\lambda_{ijk}L_{i}L_{j}E_{k}^{c}+\lambda_{ijk}^{\prime}L_{i}Q_{j}D_{k}^{c}+\frac{1}{2}\lambda_{ijk}^{\prime\prime}U_{i}^{c}D_{j}^{c}D_{k}^{c}\label{eq:superpotential}
\end{equation}

where: $\lambda_{ijk},\:\lambda_{ijk}^{\prime}$ and $\lambda_{ijk}^{\prime\prime}$
are Yukawa couplings for RPV, $i$, $j$ and $k$ are particle family
indices; $L$ and $E$ denote $SU(2)$ doublet and singlet lepton
superfields, respectively; $Q$ is a $SU(2)$ doublet quark superfield
and $U$ and $D$ are $SU(2)$ singlet quark superfields. Additionally,
it should be noted that the Yukawa couplings $\lambda_{ijk}$ and
$\lambda_{ijk}^{\prime\prime}$ are lepton number violation and baryon
number violation coefficients, respectively. 

Effective RPV interaction Lagrangians for the decays, $\widetilde{\nu}_{\tau}\rightarrow e\mu$ and $\widetilde{\nu}_{\tau}\rightarrow d \bar{d}$ are given by equations
\ref{eq:emu_lagr} and \ref{eq:mutau_lagr},
respectively.

\begin{equation}
L_{RPV_{e\mu}}=-\frac{1}{2}\lambda_{321}\widetilde{\nu}_{\tau_{L}}\overline{e}_{R}\mu_{L}-\frac{1}{2}\lambda_{312}\widetilde{\nu}_{\tau_{L}}\overline{\mu}_{R}e_{L}+h.c.\label{eq:emu_lagr}
\end{equation}

\begin{equation}
L_{RPV_{d\bar{d}}}=-\frac{1}{2}\lambda^\prime_{311}\widetilde{\nu}_{\tau_{L}}\bar{d}_{R}d_{L}+h.c.\label{eq:mutau_lagr}
\end{equation}

\begin{figure}[h]
		\scalebox{0.73}{\begin{tikzpicture}
\pgfdeclareplotmark{cross} {
\pgfpathmoveto{\pgfpoint{-0.3\pgfplotmarksize}{\pgfplotmarksize}}
\pgfpathlineto{\pgfpoint{+0.3\pgfplotmarksize}{\pgfplotmarksize}}
\pgfpathlineto{\pgfpoint{+0.3\pgfplotmarksize}{0.3\pgfplotmarksize}}
\pgfpathlineto{\pgfpoint{+1\pgfplotmarksize}{0.3\pgfplotmarksize}}
\pgfpathlineto{\pgfpoint{+1\pgfplotmarksize}{-0.3\pgfplotmarksize}}
\pgfpathlineto{\pgfpoint{+0.3\pgfplotmarksize}{-0.3\pgfplotmarksize}}
\pgfpathlineto{\pgfpoint{+0.3\pgfplotmarksize}{-1.\pgfplotmarksize}}
\pgfpathlineto{\pgfpoint{-0.3\pgfplotmarksize}{-1.\pgfplotmarksize}}
\pgfpathlineto{\pgfpoint{-0.3\pgfplotmarksize}{-0.3\pgfplotmarksize}}
\pgfpathlineto{\pgfpoint{-1.\pgfplotmarksize}{-0.3\pgfplotmarksize}}
\pgfpathlineto{\pgfpoint{-1.\pgfplotmarksize}{0.3\pgfplotmarksize}}
\pgfpathlineto{\pgfpoint{-0.3\pgfplotmarksize}{0.3\pgfplotmarksize}}
\pgfpathclose
\pgfusepathqstroke
}
\pgfdeclareplotmark{cross*} {
\pgfpathmoveto{\pgfpoint{-0.3\pgfplotmarksize}{\pgfplotmarksize}}
\pgfpathlineto{\pgfpoint{+0.3\pgfplotmarksize}{\pgfplotmarksize}}
\pgfpathlineto{\pgfpoint{+0.3\pgfplotmarksize}{0.3\pgfplotmarksize}}
\pgfpathlineto{\pgfpoint{+1\pgfplotmarksize}{0.3\pgfplotmarksize}}
\pgfpathlineto{\pgfpoint{+1\pgfplotmarksize}{-0.3\pgfplotmarksize}}
\pgfpathlineto{\pgfpoint{+0.3\pgfplotmarksize}{-0.3\pgfplotmarksize}}
\pgfpathlineto{\pgfpoint{+0.3\pgfplotmarksize}{-1.\pgfplotmarksize}}
\pgfpathlineto{\pgfpoint{-0.3\pgfplotmarksize}{-1.\pgfplotmarksize}}
\pgfpathlineto{\pgfpoint{-0.3\pgfplotmarksize}{-0.3\pgfplotmarksize}}
\pgfpathlineto{\pgfpoint{-1.\pgfplotmarksize}{-0.3\pgfplotmarksize}}
\pgfpathlineto{\pgfpoint{-1.\pgfplotmarksize}{0.3\pgfplotmarksize}}
\pgfpathlineto{\pgfpoint{-0.3\pgfplotmarksize}{0.3\pgfplotmarksize}}
\pgfpathclose
\pgfusepathqfillstroke
}
\pgfdeclareplotmark{newstar} {
\pgfpathmoveto{\pgfqpoint{0pt}{\pgfplotmarksize}}
\pgfpathlineto{\pgfqpointpolar{44}{0.5\pgfplotmarksize}}
\pgfpathlineto{\pgfqpointpolar{18}{\pgfplotmarksize}}
\pgfpathlineto{\pgfqpointpolar{-20}{0.5\pgfplotmarksize}}
\pgfpathlineto{\pgfqpointpolar{-54}{\pgfplotmarksize}}
\pgfpathlineto{\pgfqpointpolar{-90}{0.5\pgfplotmarksize}}
\pgfpathlineto{\pgfqpointpolar{234}{\pgfplotmarksize}}
\pgfpathlineto{\pgfqpointpolar{198}{0.5\pgfplotmarksize}}
\pgfpathlineto{\pgfqpointpolar{162}{\pgfplotmarksize}}
\pgfpathlineto{\pgfqpointpolar{134}{0.5\pgfplotmarksize}}
\pgfpathclose
\pgfusepathqstroke
}
\pgfdeclareplotmark{newstar*} {
\pgfpathmoveto{\pgfqpoint{0pt}{\pgfplotmarksize}}
\pgfpathlineto{\pgfqpointpolar{44}{0.5\pgfplotmarksize}}
\pgfpathlineto{\pgfqpointpolar{18}{\pgfplotmarksize}}
\pgfpathlineto{\pgfqpointpolar{-20}{0.5\pgfplotmarksize}}
\pgfpathlineto{\pgfqpointpolar{-54}{\pgfplotmarksize}}
\pgfpathlineto{\pgfqpointpolar{-90}{0.5\pgfplotmarksize}}
\pgfpathlineto{\pgfqpointpolar{234}{\pgfplotmarksize}}
\pgfpathlineto{\pgfqpointpolar{198}{0.5\pgfplotmarksize}}
\pgfpathlineto{\pgfqpointpolar{162}{\pgfplotmarksize}}
\pgfpathlineto{\pgfqpointpolar{134}{0.5\pgfplotmarksize}}
\pgfpathclose
\pgfusepathqfillstroke
}
\definecolor{c}{rgb}{1,1,1};
\draw [color=c, fill=c] (2,1.36103) rectangle (18,12.2493);
\definecolor{c}{rgb}{0,0,0};
\draw [c,line width=0.9] (2,1.36103) -- (2,12.2493) -- (18,12.2493) -- (18,1.36103) -- (2,1.36103);
\definecolor{c}{rgb}{1,1,1};
\draw [color=c, fill=c] (2,1.36103) rectangle (18,12.2493);
\definecolor{c}{rgb}{0,0,0};
\draw [c,line width=0.9] (2,1.36103) -- (2,12.2493) -- (18,12.2493) -- (18,1.36103) -- (2,1.36103);
\draw [c,line width=0.9] (2,1.36103) -- (18,1.36103);
\draw [c,line width=0.9] (2.69945,1.68768) -- (2.69945,1.36103);
\draw [c,line width=0.9] (4.88525,1.68768) -- (4.88525,1.36103);
\draw [c,line width=0.9] (7.07104,1.68768) -- (7.07104,1.36103);
\draw [c,line width=0.9] (9.25683,1.68768) -- (9.25683,1.36103);
\draw [c,line width=0.9] (11.4426,1.68768) -- (11.4426,1.36103);
\draw [c,line width=0.9] (13.6284,1.68768) -- (13.6284,1.36103);
\draw [c,line width=0.9] (15.8142,1.68768) -- (15.8142,1.36103);
\draw [c,line width=0.9] (18,1.68768) -- (18,1.36103);
\draw [c,line width=0.9] (2.69945,1.68768) -- (2.69945,1.36103);
\draw [anchor=base] (2.69945,0.911891) node[scale=1.08185, color=c, rotate=0]{5};
\draw [anchor=base] (4.88525,0.911891) node[scale=1.08185, color=c, rotate=0]{10};
\draw [anchor=base] (7.07104,0.911891) node[scale=1.08185, color=c, rotate=0]{15};
\draw [anchor=base] (9.25683,0.911891) node[scale=1.08185, color=c, rotate=0]{20};
\draw [anchor=base] (11.4426,0.911891) node[scale=1.08185, color=c, rotate=0]{25};
\draw [anchor=base] (13.6284,0.911891) node[scale=1.08185, color=c, rotate=0]{30};
\draw [anchor=base] (15.8142,0.911891) node[scale=1.08185, color=c, rotate=0]{35};
\draw [anchor=base] (18,0.911891) node[scale=1.08185, color=c, rotate=0]{40};
\draw (10,0.446418) node[scale=1.58185, color=c, rotate=0]{M$_{\tilde{\nu_{\tau}}}$ [TeV]};
\draw [c,line width=0.9] (2,12.2493) -- (18,12.2493);
\draw [c,line width=0.9] (2.69945,11.9226) -- (2.69945,12.2493);
\draw [c,line width=0.9] (4.88525,11.9226) -- (4.88525,12.2493);
\draw [c,line width=0.9] (7.07104,11.9226) -- (7.07104,12.2493);
\draw [c,line width=0.9] (9.25683,11.9226) -- (9.25683,12.2493);
\draw [c,line width=0.9] (11.4426,11.9226) -- (11.4426,12.2493);
\draw [c,line width=0.9] (13.6284,11.9226) -- (13.6284,12.2493);
\draw [c,line width=0.9] (15.8142,11.9226) -- (15.8142,12.2493);
\draw [c,line width=0.9] (18,11.9226) -- (18,12.2493);
\draw [c,line width=0.9] (2.69945,11.9226) -- (2.69945,12.2493);
\draw [c,line width=0.9] (2,1.36103) -- (2,12.2493);
\draw [c,line width=0.9] (2.24,1.86405) -- (2,1.86405);
\draw [c,line width=0.9] (2.48,2.31402) -- (2,2.31402);
\draw [anchor= east] (1.844,2.31402) node[scale=1.08185, color=c, rotate=0]{10};
\draw [c,line width=0.9] (2.24,5.2743) -- (2,5.2743);
\draw [c,line width=0.9] (2.24,7.00595) -- (2,7.00595);
\draw [c,line width=0.9] (2.24,8.23458) -- (2,8.23458);
\draw [c,line width=0.9] (2.24,9.18758) -- (2,9.18758);
\draw [c,line width=0.9] (2.24,9.96623) -- (2,9.96623);
\draw [c,line width=0.9] (2.24,10.6246) -- (2,10.6246);
\draw [c,line width=0.9] (2.24,11.1949) -- (2,11.1949);
\draw [c,line width=0.9] (2.24,11.6979) -- (2,11.6979);
\draw [c,line width=0.9] (2.48,12.1479) -- (2,12.1479);
\draw [anchor= east] (1.844,12.1479) node[scale=1.08185, color=c, rotate=0]{$10^{2}$};
\draw (0.88,6.80516) node[scale=1.58185, color=c, rotate=90]{$\Gamma [GeV]$};
\draw [c,line width=0.9] (18,1.36103) -- (18,12.2493);
\draw [c,line width=0.9] (17.76,1.86405) -- (18,1.86405);
\draw [c,line width=0.9] (17.52,2.31402) -- (18,2.31402);
\draw [c,line width=0.9] (17.76,5.2743) -- (18,5.2743);
\draw [c,line width=0.9] (17.76,7.00595) -- (18,7.00595);
\draw [c,line width=0.9] (17.76,8.23458) -- (18,8.23458);
\draw [c,line width=0.9] (17.76,9.18758) -- (18,9.18758);
\draw [c,line width=0.9] (17.76,9.96623) -- (18,9.96623);
\draw [c,line width=0.9] (17.76,10.6246) -- (18,10.6246);
\draw [c,line width=0.9] (17.76,11.1949) -- (18,11.1949);
\draw [c,line width=0.9] (17.76,11.6979) -- (18,11.6979);
\draw [c,line width=0.9] (17.52,12.1479) -- (18,12.1479);
\definecolor{c}{rgb}{0,0,1};
\draw [c,line width=3.6] (2.01069,1.36103) -- (2.04372,1.46051);
\draw [c,line width=3.6] (2.04372,1.46051) -- (2.2623,2.03079) -- (2.48087,2.53381) -- (2.69945,2.98379) -- (2.91803,3.39083) -- (3.13661,3.76244) -- (3.35519,4.10428) -- (3.57377,4.42078) -- (3.79235,4.71544) -- (4.01093,4.99107) --
 (4.22951,5.24998) -- (4.44809,5.49409) -- (4.66667,5.725) -- (4.88525,5.94406) -- (5.10383,6.15243) -- (5.3224,6.35111) -- (5.54098,6.54095) -- (5.75956,6.72272) -- (5.97814,6.89706) -- (6.19672,7.06456) -- (6.4153,7.22574) -- (6.63388,7.38106) --
 (6.85246,7.53093) -- (7.07104,7.67571) -- (7.28962,7.81575) -- (7.5082,7.95134) -- (7.72678,8.08276) -- (7.94536,8.21026) -- (8.16393,8.33406) -- (8.38251,8.45437) -- (8.60109,8.57138) -- (8.81967,8.68528) -- (9.03825,8.79621) -- (9.25683,8.90434)
 -- (9.47541,9.0098) -- (9.69399,9.11271) -- (9.91257,9.21321) -- (10.1311,9.31139) -- (10.3497,9.40737) -- (10.5683,9.50123) -- (10.7869,9.59308) -- (11.0055,9.68299) -- (11.224,9.77106) -- (11.4426,9.85734) -- (11.6612,9.94191) -- (11.8798,10.0248)
 -- (12.0984,10.1062) -- (12.3169,10.186) -- (12.5355,10.2644) -- (12.7541,10.3413) -- (12.9727,10.4169) -- (13.1913,10.4912) -- (13.4098,10.5642) -- (13.6284,10.636) -- (14.0656,10.776) -- (14.5027,10.9116) -- (14.9399,11.043) -- (15.377,11.1705) --
 (15.8142,11.2943) -- (16.2514,11.4146) -- (16.6885,11.5317) -- (17.1257,11.6456) -- (17.5628,11.7565) -- (18,11.8646);
\definecolor{c}{rgb}{0,0,0};
\draw [anchor=base west] (4.25,10.582) node[scale=1.90914, color=c, rotate=0]{$\lambda_{ijk}$  = 0.07};
\end{tikzpicture}}
\caption{Total decay width channels for R-parity violated interactions. \label{fig:Decay-width}}
\end{figure}
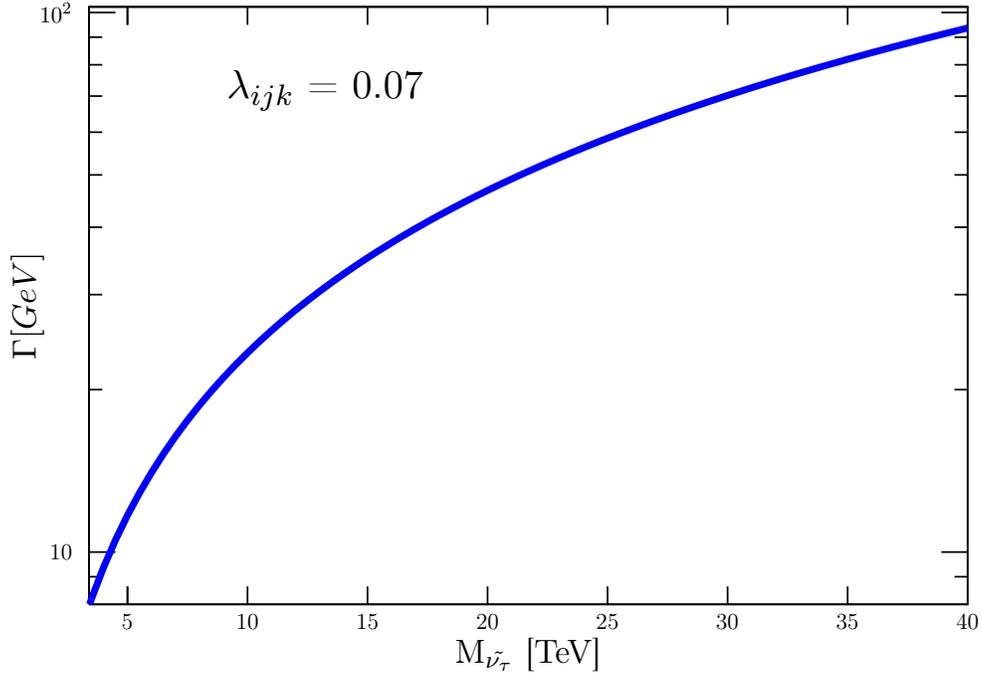

\begin{figure}[h]
\includegraphics[scale=0.6]{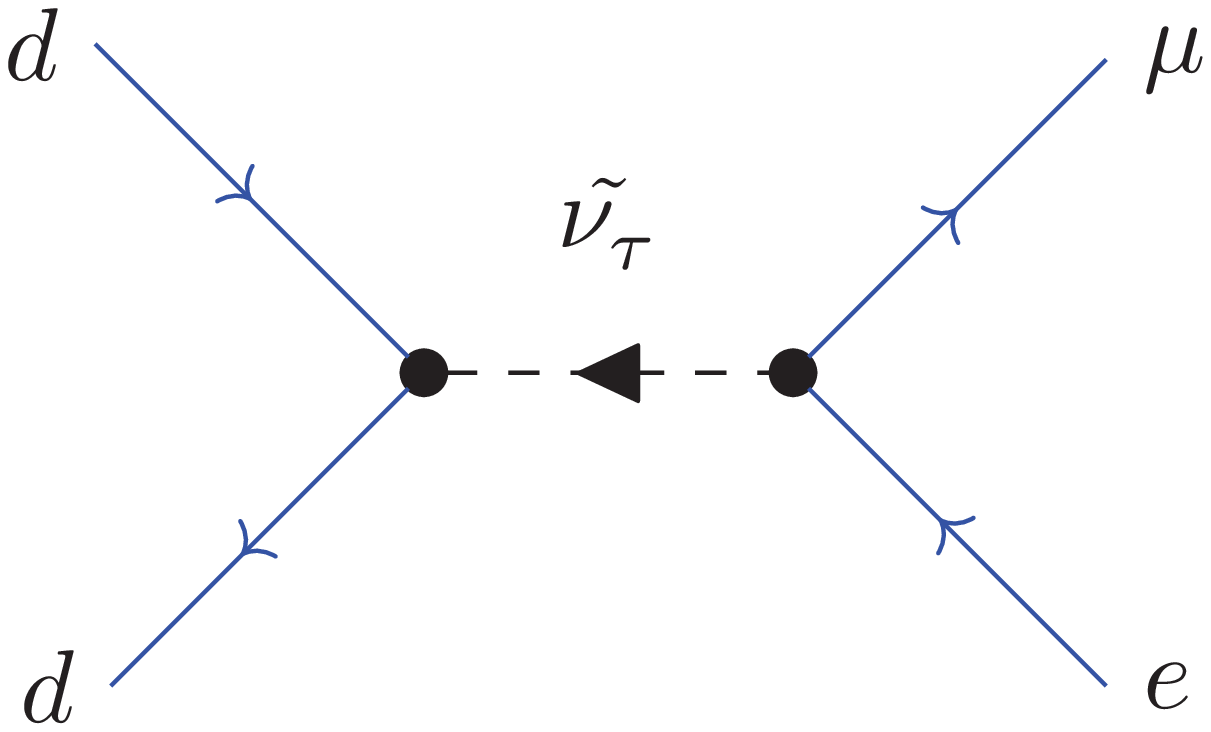}\includegraphics[scale=0.6]{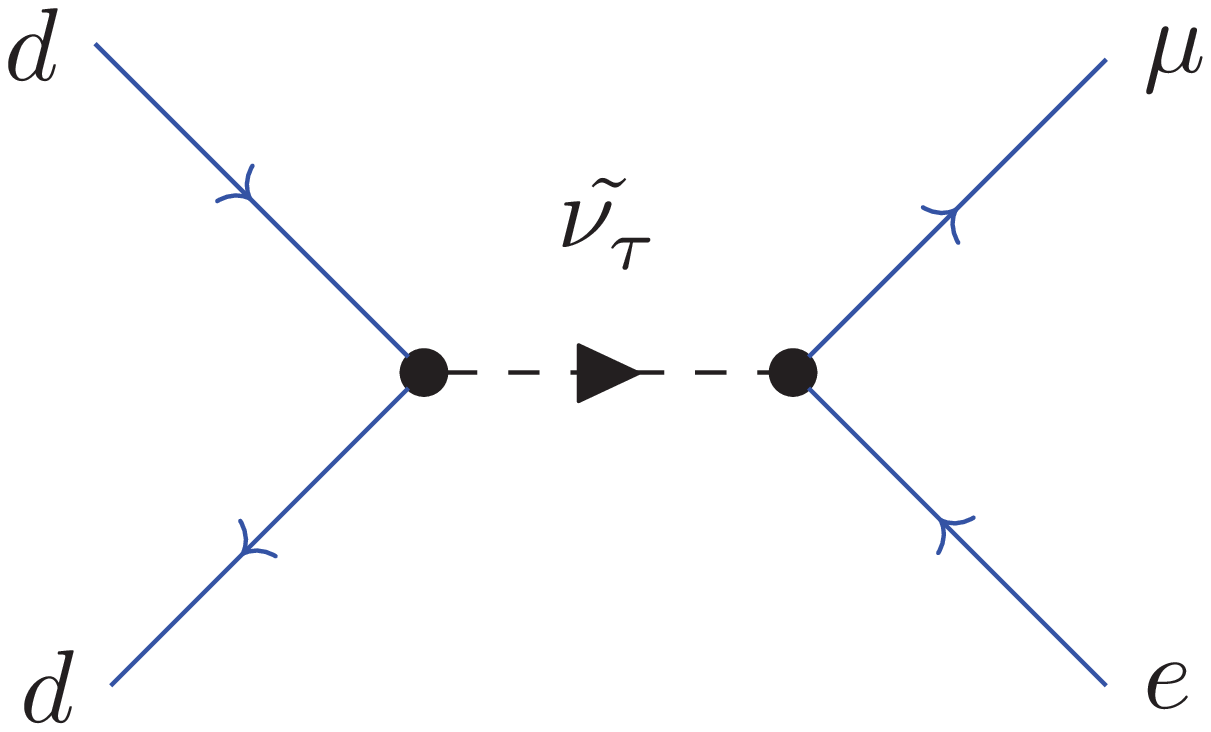}
\caption{Feynman diagrams of the $d\bar{d}\rightarrow e\mu$ subprocess. \label{fig:Feynman-diagrams}}
\end{figure}

When decay width is calculated, all RPV decay channels are considered.
Then, total decay widths with respect to $\widetilde{\nu}_{\tau}$
mass was plotted in Fig. \ref{fig:Decay-width}. Resonant production
of the $\widetilde{\nu}_{\tau}$ at FCC via RPV interactions is illustrated
in Fig. \ref{fig:Feynman-diagrams} as Feynman diagrams. Undoubtedly,
black dots represent the R-partity violation vertices. 

\begin{figure}[h]
\scalebox{0.73}{\input{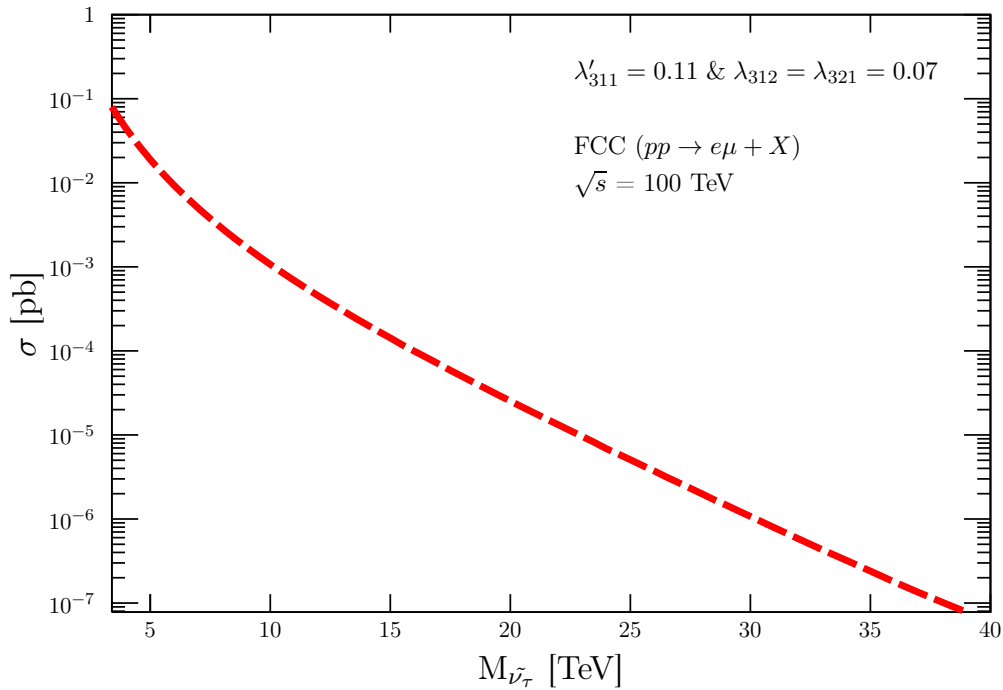}}
\caption{Resonant production cross section of $\widetilde{\nu}_{\tau}$ in the FCC-pp at $\sqrt{{s}}=100$ TeV. \label{fig:Cross-section}}

\end{figure}

This work considers resonant $\widetilde{\nu}_{\tau}$ generation
via $\lambda_{311}^{\prime}$ coupling at FCC-pp collider and $\widetilde{\nu}_{\tau}$
decays into $e\mu$ final states via $\lambda_{312}=\lambda_{321}$
couplings. So, signal processes, $d\bar{d}\rightarrow\widetilde{\nu}_{\tau}+X\rightarrow e\mu+X$,
were produced at leading order (LO) with MadGraph5\_aMC@NLOv2.3.3
\citep{Alwall2014,FUKS_2012} for numerical calculation, parton distribution
function (PDF) was set as CTEQ6L1 \citep{pumplin2002,stump2003},
renormalization and factorization scales were identified as $\widetilde{\nu}_{\tau}$
mass, and $\lambda_{311}^{\prime}=0.11$ and $\lambda_{312}=\lambda_{321}=0.07$
were taken as RPV Yukawa couplings. One can see from Fig. \ref{fig:Cross-section}
that $\widetilde{\nu}_{\tau}$ could be produced up to roughly 30
TeV mass value when considering 10 events with 17500 $fb^{-1}$ luminosity
value at the FCC-pp. 

\section{Signal - Background Analysis\label{sec:Signal---Background}}

As it is mentioned in previous section, the signal processes were
generated as $d\bar{d}\rightarrow\widetilde{\nu}_{\tau}+X\rightarrow e\mu+X$
with analysis software, MadGraph5\_aMC@NLOv2.3.3. The background processes
were comprised of two categories: irreducible and reducible backgrounds
for $e\mu$ final states. Irreducible background is composed of final
state particles with two different lepton flavors. Top pair production
($t\bar{t}$), single top ($tW$) and diboson ($WW,\:WZ,\:ZZ$) contribute
to irreducible background. Consequence of mis-reconstruction of jets
as lepton, $W$+jets and multi-jets and mis-reconstruction of photon
as lepton, $W+\gamma$, the reducible background occurs which is neglected
due to its small contribution to total background \citep{Aad:2014pda,Aaboud2016,ALVAREZ201719}
for $e\mu$ final states. Numerical calculation of background and
signal processes were computed by MadGraph5\_aMC@NLOv2.3.3. 

\begin{figure}[h]
\includegraphics[scale=0.42]{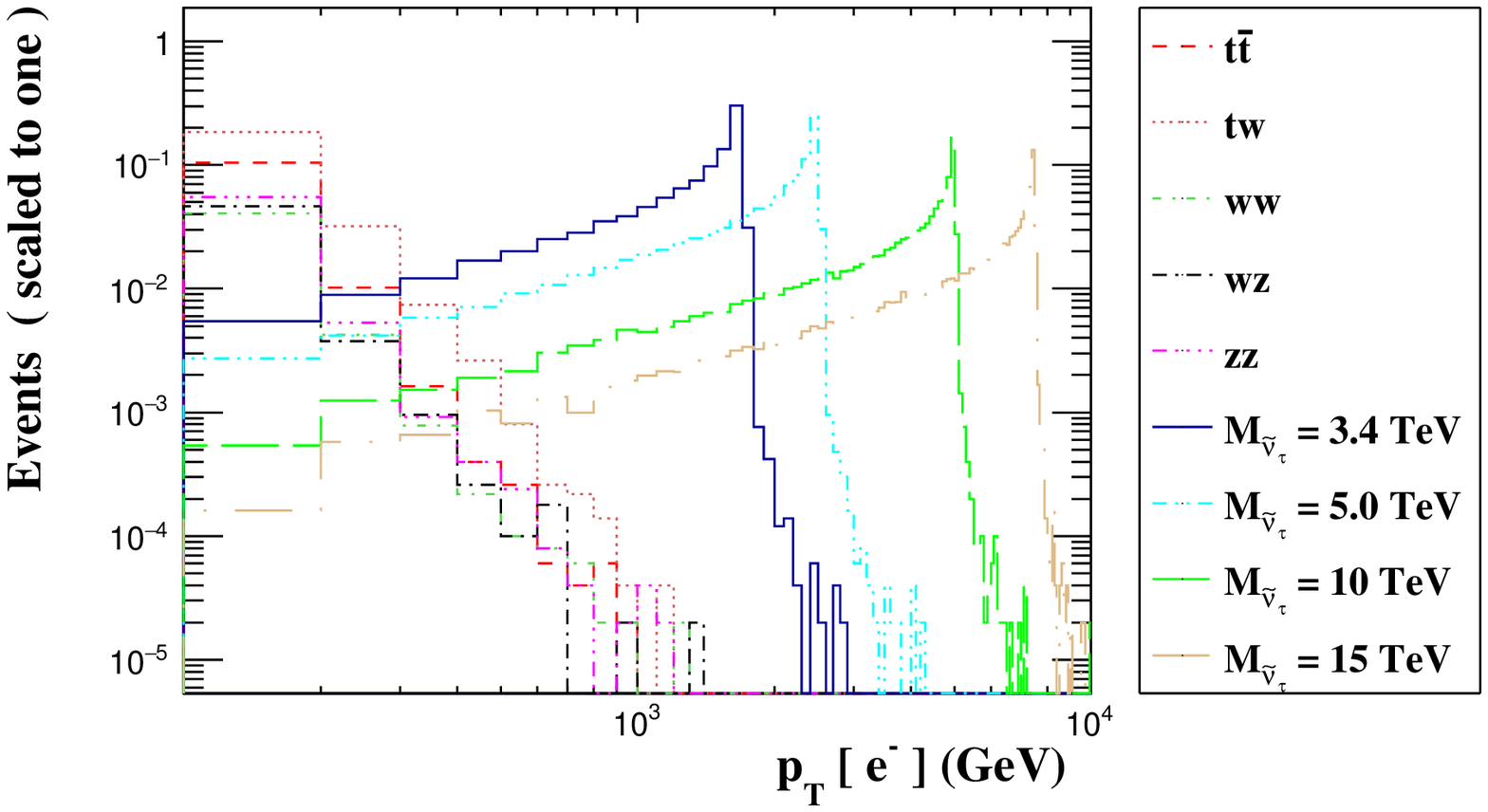}\includegraphics[scale=0.42]{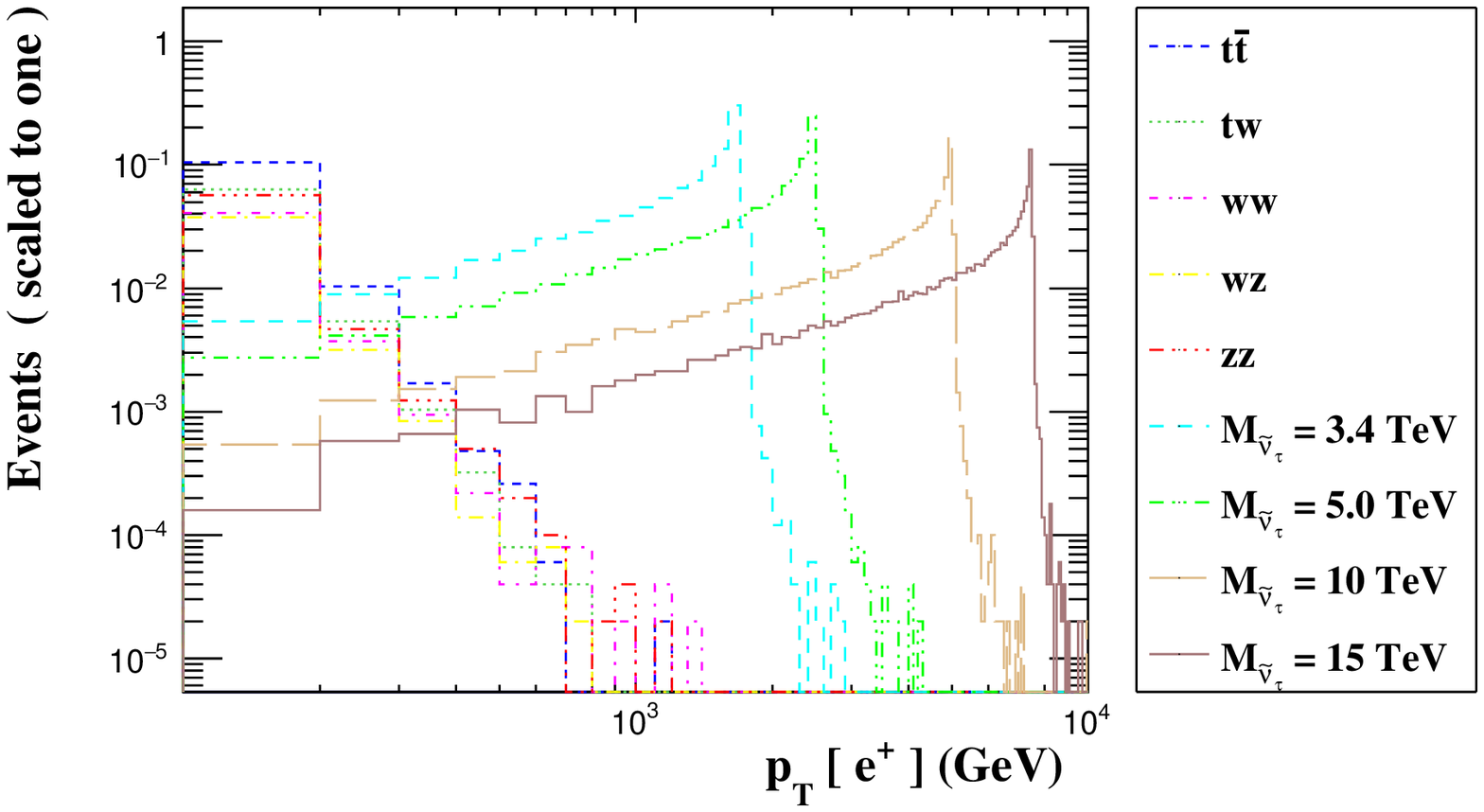}
\includegraphics[scale=0.42]{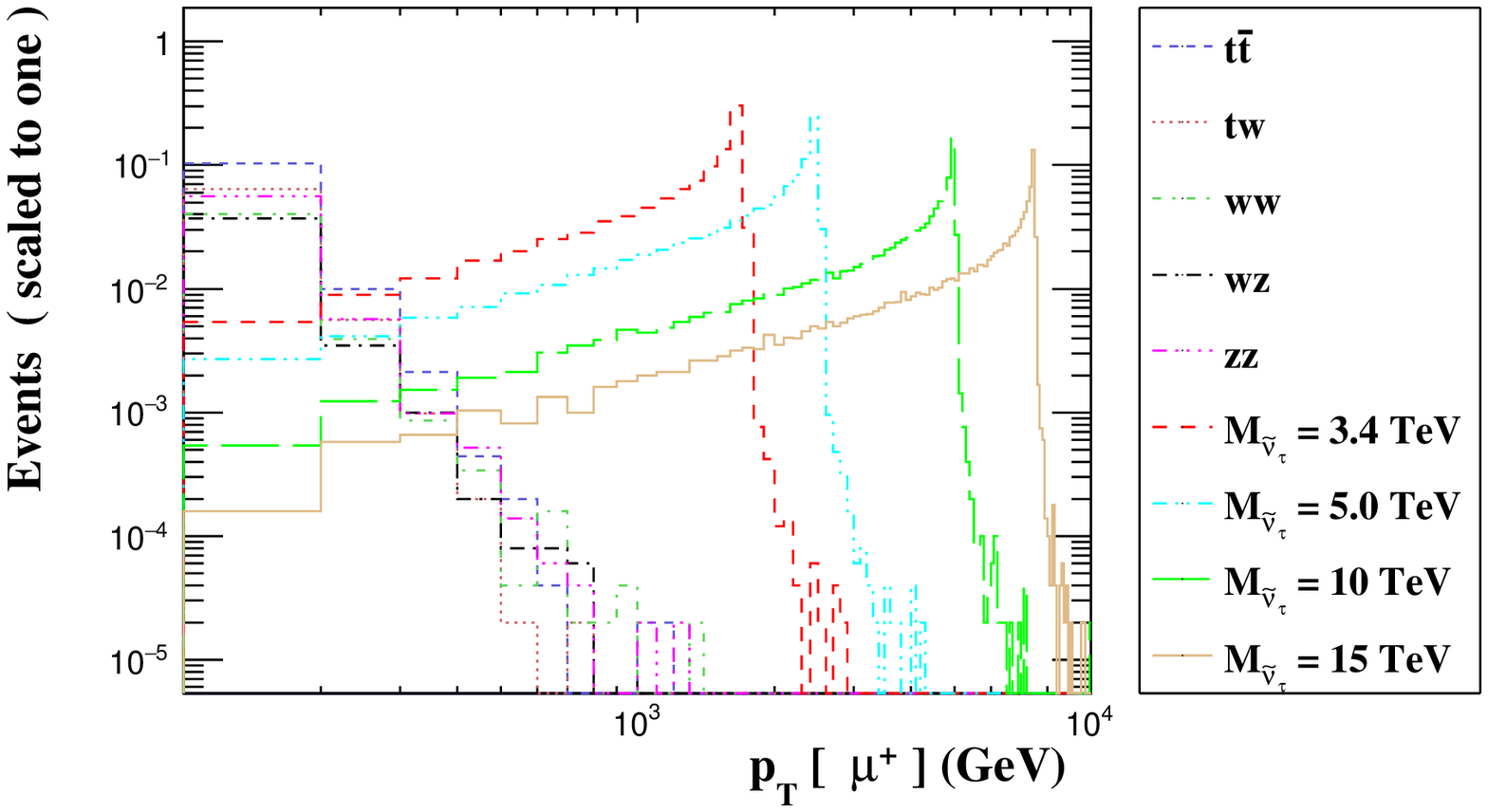}\includegraphics[scale=0.42]{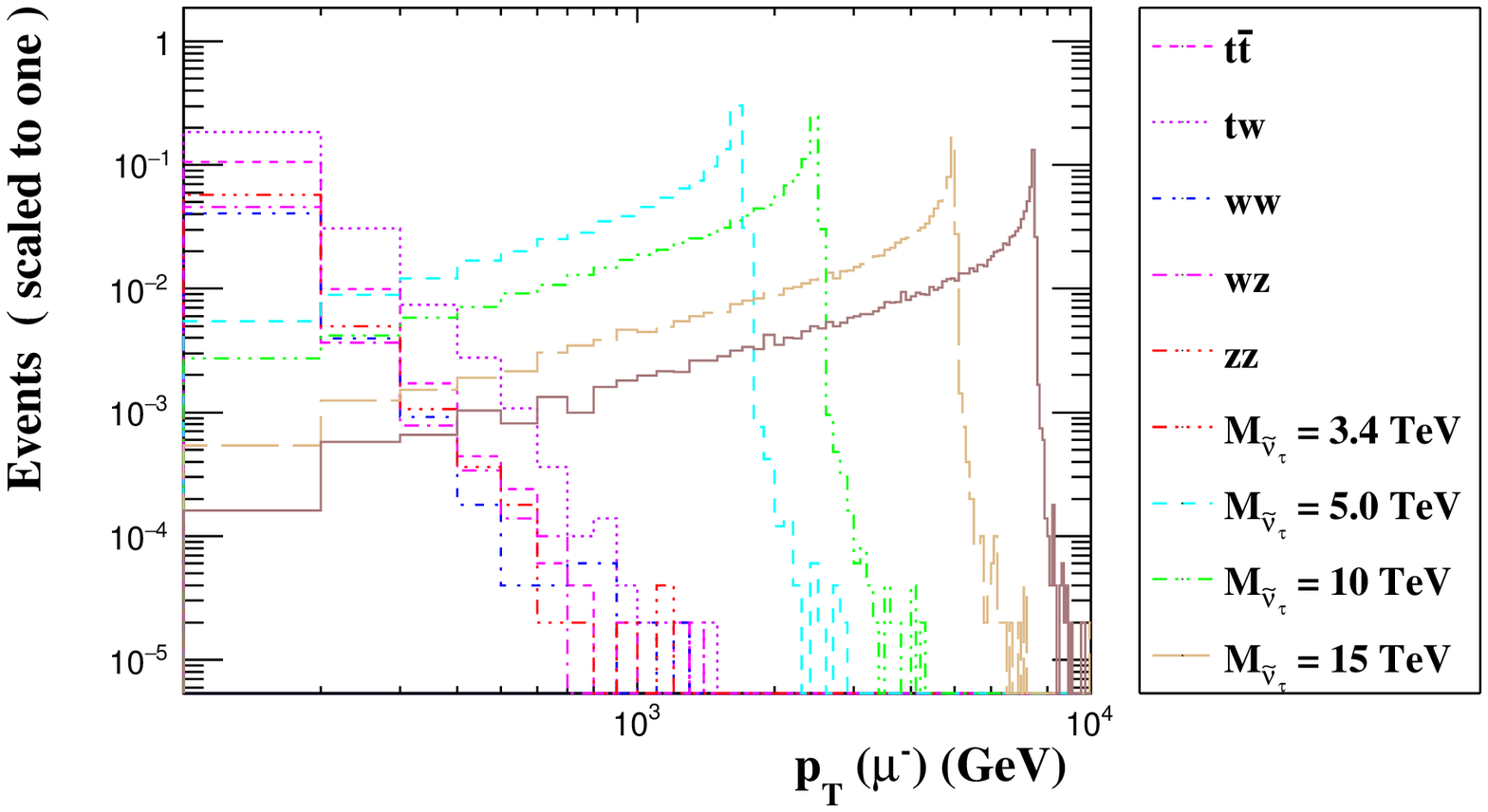}
\caption{Transverse momentum plots for electron, positron, anti-muon and muon.\label{fig:Transverse-momentum-plots}}
\end{figure}

\begin{figure}[h]
\includegraphics[scale=0.42]{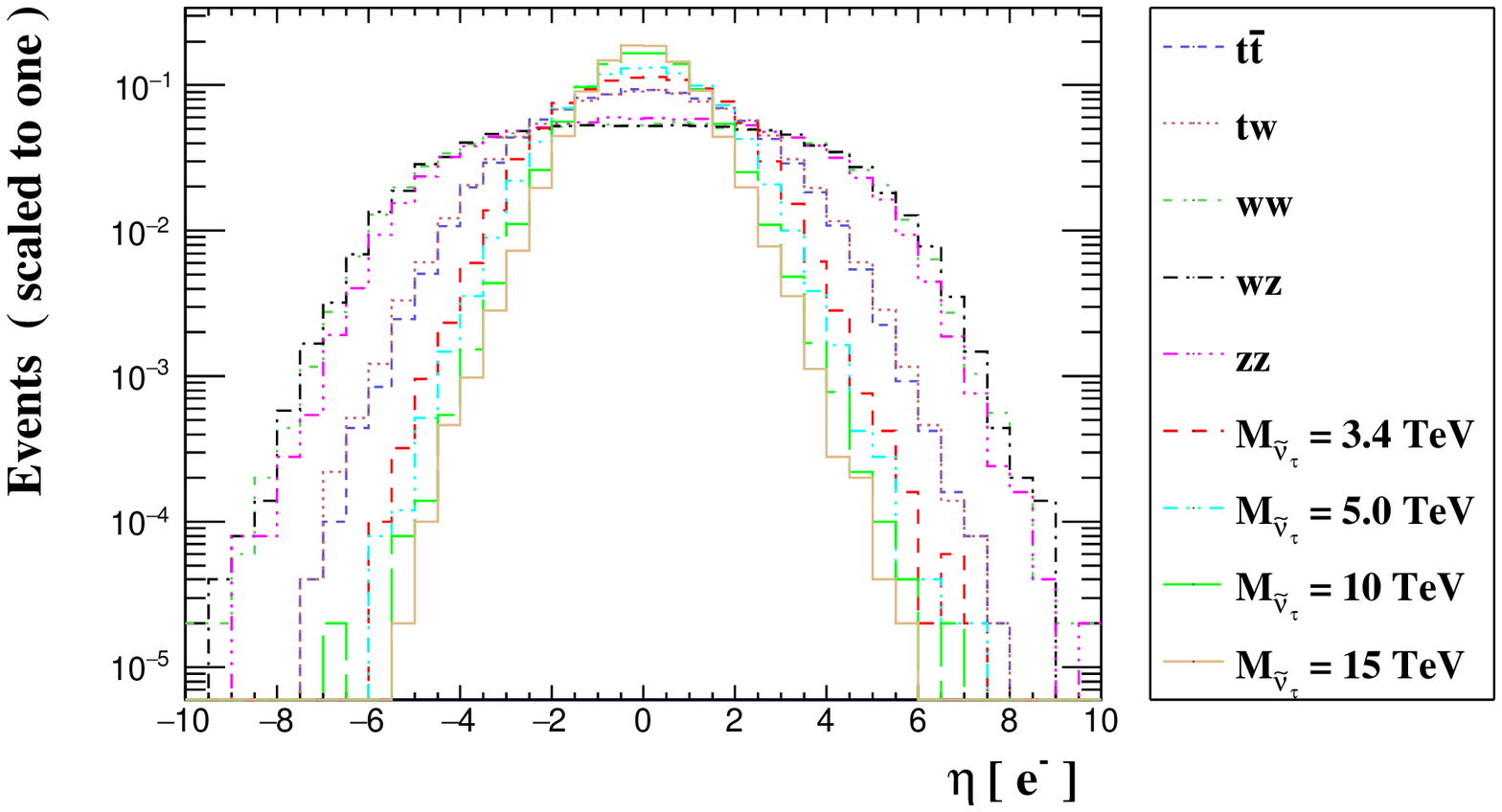}\includegraphics[scale=0.42]{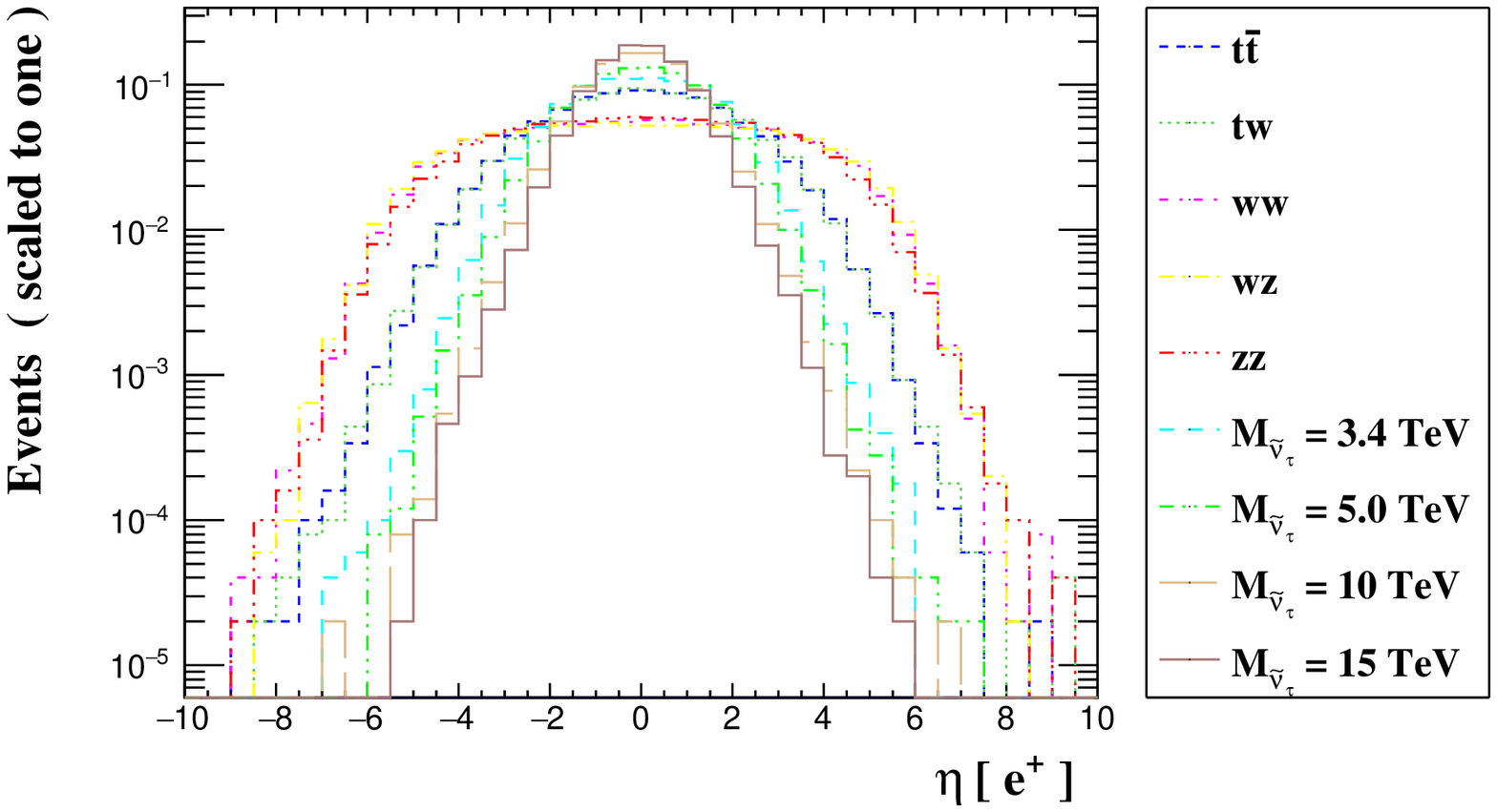}
\includegraphics[scale=0.42]{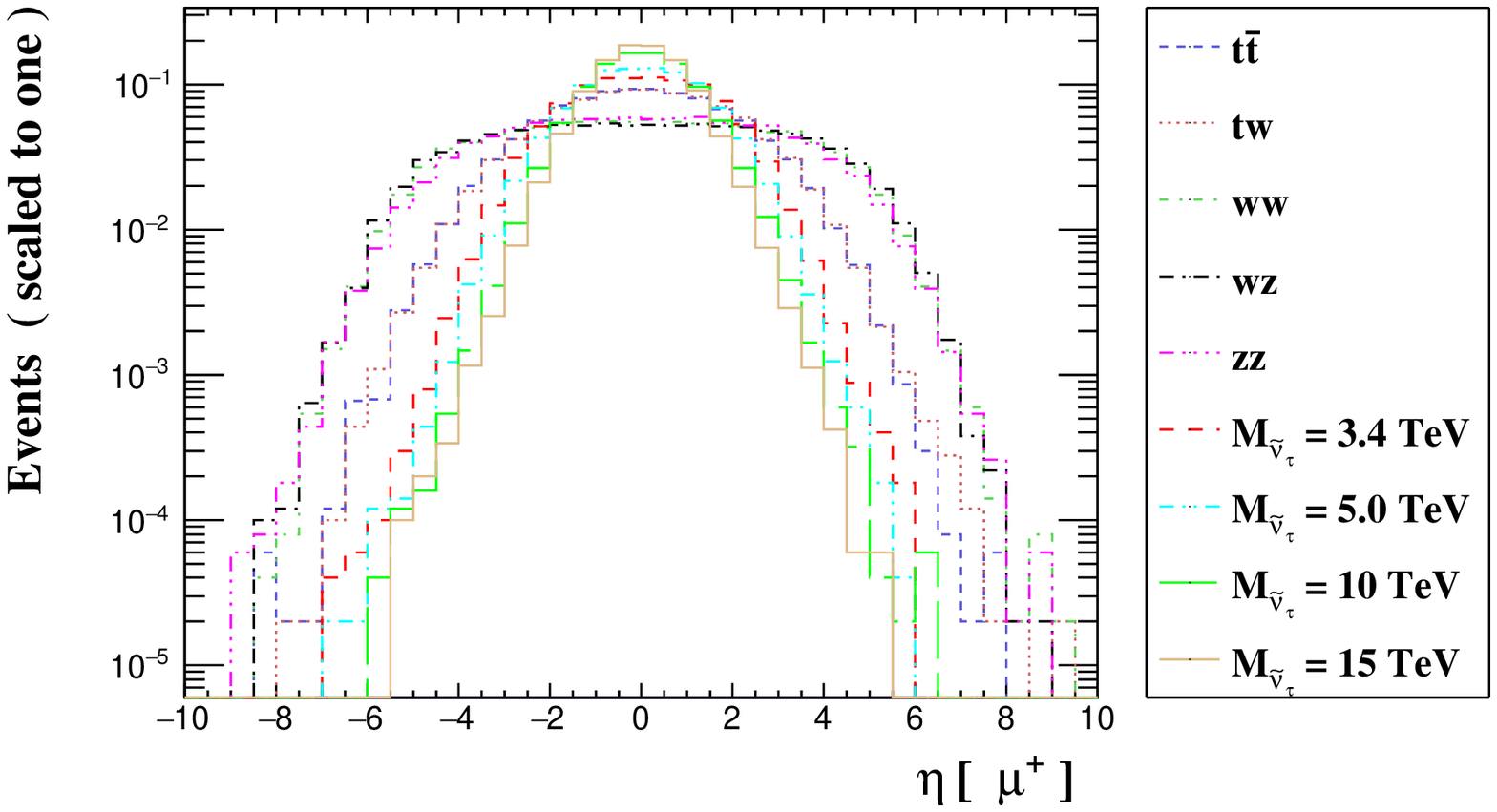}\includegraphics[scale=0.42]{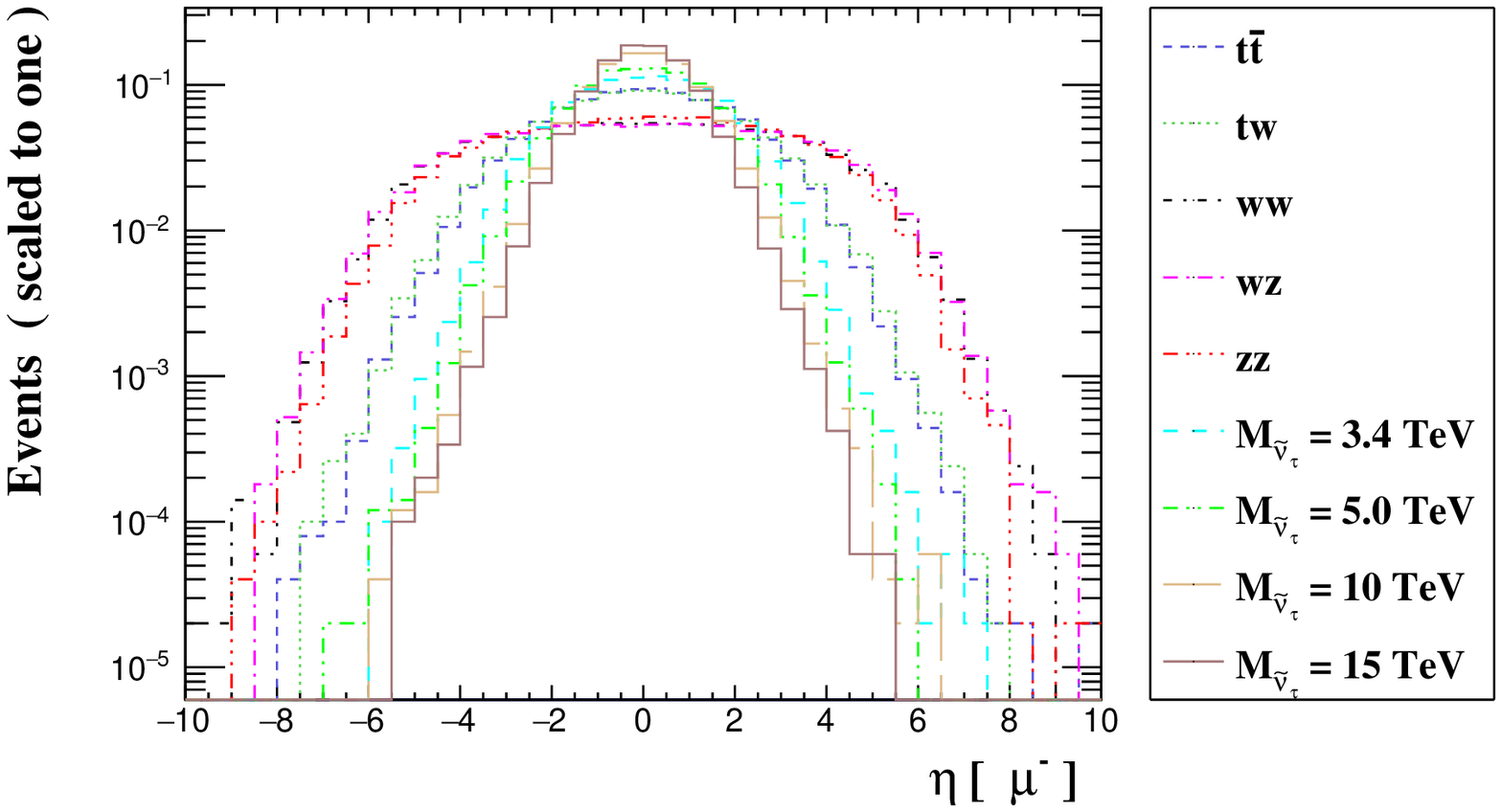}
\caption{Pseudo rapidity plots for electron, positron, anti-muon and muon.\label{fig:rapidity-plots}}
\end{figure}

\begin{figure}[h]
\scalebox{0.83}{\input{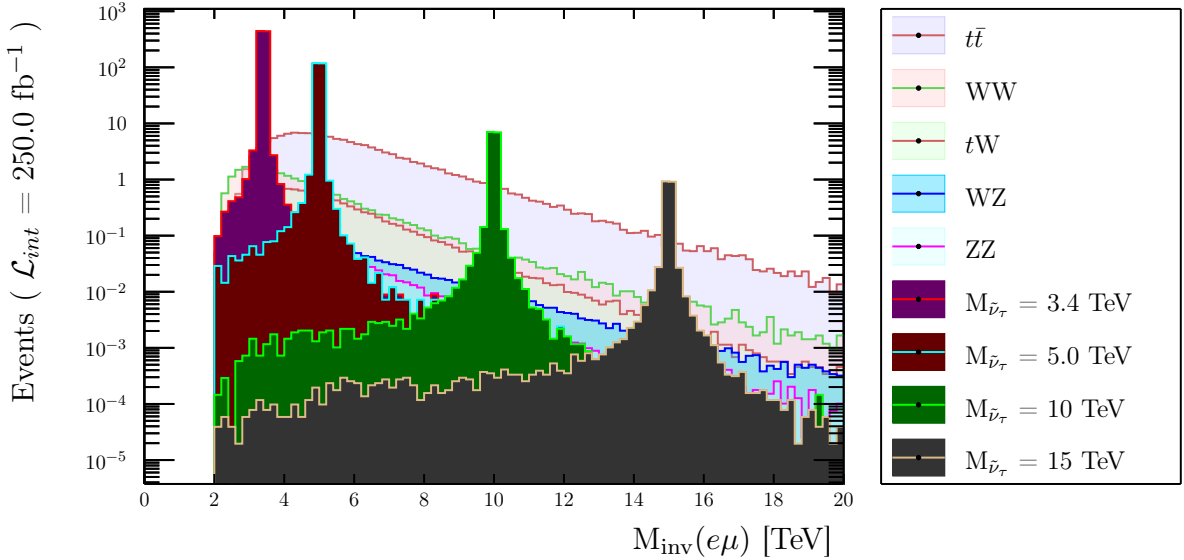}}
\caption{Signal and background processes invariant mass distributions for $e\mu$ final states with 1 TeV transverse momentum cut for final leptons. \label{fig:Invariant-mass-plot}}

\end{figure}

Transverse momentum ($P_{T}$), pseudorapidity ($\eta$) and invariant
mass ($M_{\widetilde{\nu}_{\tau}}$) distribution plots for $e\mu$
final states were produced to identify appropriate cuts that are created
by MADANALYSIS 5 \citep{CONTE2013222}. It is seen from Fig. \ref{fig:Transverse-momentum-plots}
that selecting 1000 GeV transverse momentum cut for $e\mu$ final
state particles is enough to make signal almost unchanged and to reduce
background. Also, Fig. \ref{fig:rapidity-plots} shows that signal
distributions of the $e\mu$ final state particles exceed the background between -2.5
and 2.5 $\eta$ values which coincide with CMS and ATLAS detector's acceptances \citep{Aaboud:2018, Sirunyan2018}. 
For this reason, $|\eta| < $ 2.5 region is chosen for analysis calculations. Then, as can be seen from Fig. \ref{fig:Invariant-mass-plot}, selecting
$M_{\widetilde{\nu}_{\tau}}-1000<M_{\widetilde{\nu}_{\tau}}<M_{\widetilde{\nu}_{\tau}}+1000$
mass window is a proper cut to eliminate background effects. Additionally,
$E_{T}^{miss}<25$ GeV are taken to suppress the first two generation
neutrinos' contribution to background that come from $W$ decays.
Lastly, $\Delta R\geq0.4$ cone angle cut is applied to $e\mu$ final
states.

\begin{figure}[h]
\scalebox{0.9}{\input{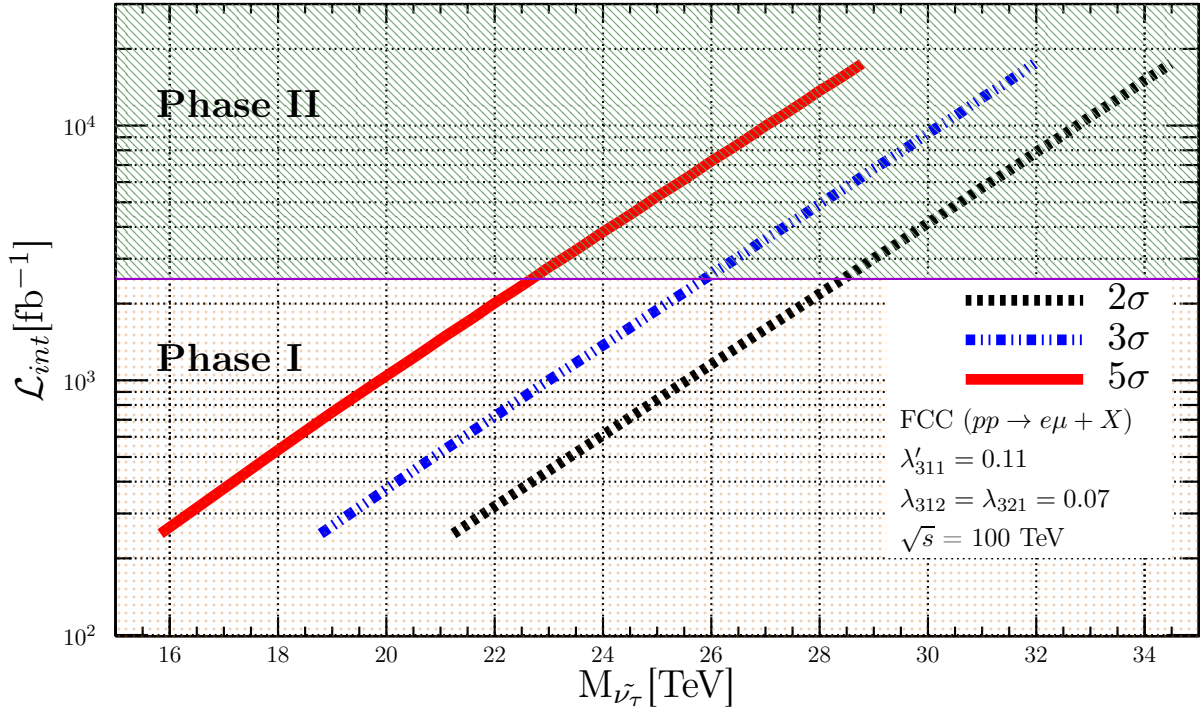}}	
\caption{Scalar tau neutrino mass limits at FCC-pp for both phase's luminosity values. \label{fig:lumiMass}}
\end{figure}

Eq. \ref{eq:significance} was used for calculating statistical significances;

\begin{equation}
SS=\frac{\sigma_{S}}{\sqrt{{\sigma_{s}+\sigma_{B}}}}\sqrt{{\mathcal{L}_{int}}}\label{eq:significance}
\end{equation}

where, $\sigma_{S}$ and $\sigma_{B}$ denote signal and background
cross section values, respectively and $\mathcal{L}_{int}$ is an
integrated luminosity value. Using Phase I and II luminosity values
for 25 years operation time and Eq. \ref{eq:significance} by applying
cuts mentioned above, $\widetilde{\nu}_{\tau}$ mass limits depending
on luminosity was plotted in Fig. \ref{fig:lumiMass} for three confidence
levels. As can be seen from Fig. \ref{fig:lumiMass}, $\widetilde{\nu}_{\tau}$
will be discovered up to nearly 29 TeV mass value for integrated luminosity
equals to 17500 $fb^{-1}$ with taking $\lambda_{311}^{\prime}=0.11$
and $\lambda_{312}=\lambda_{321}=0.07$. 

\begin{figure}[h]
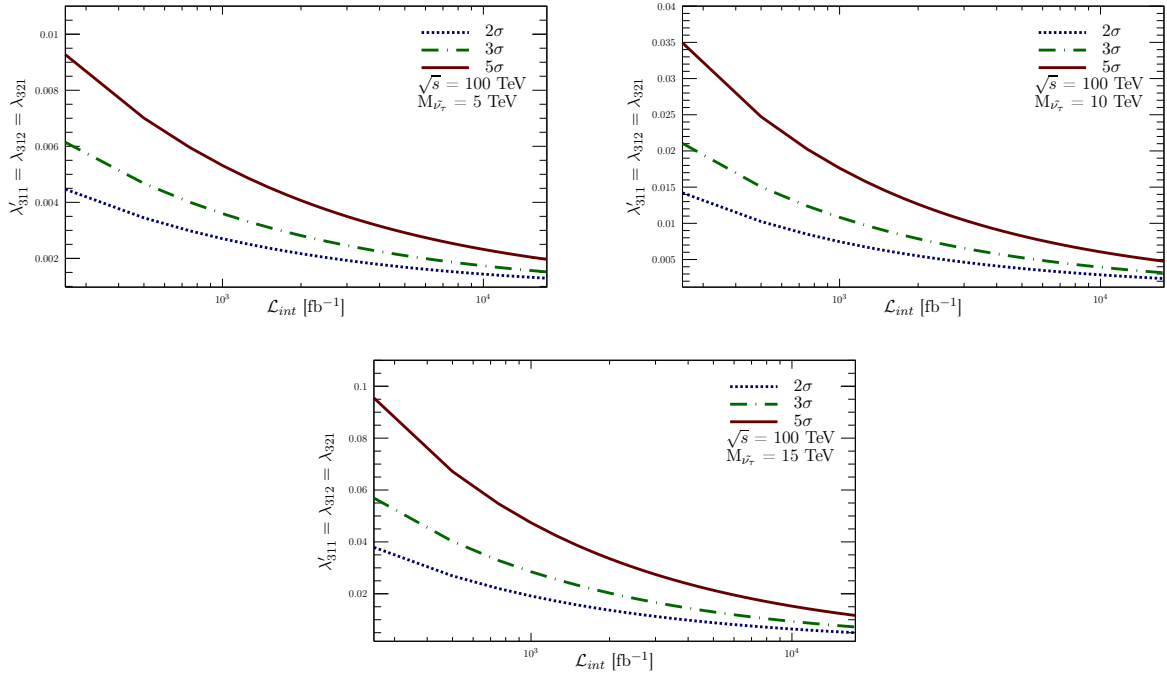

\scalebox{0.4}{\input{LumiLamdaCoeff5TeV.tex}}	\scalebox{0.4}{\input{LumiLamdaCoeff10TeV.tex}}	\scalebox{0.4}{\input{LumiLamdaCoeff15TeV.tex}}	
\caption{Yukawa coupling constants limits for 5, 10 and 15 TeV scalar tau neutrino masses. \label{fig:ucu-yanyana.}}
\end{figure}

Yukawa coupling coefficients, $\lambda_{311}^{\prime}$, $\lambda_{312}$
and $\lambda_{321}$ can take different values other than 0.11 and
0.07. In numerical calculations, those values were considered but
taking equal all $\lambda'_{311}$, $\lambda_{312}$ and $\lambda_{321}$
each other, for 5, 10 and 15 TeV mass values of the $\widetilde{\nu}_{\tau}$,
the coupling coefficients were scanned with whole integrated luminosities.
Fig. \ref{fig:ucu-yanyana.} shows that if $\widetilde{\nu}_{\tau}$
mass value is 5 TeV, $\widetilde{\nu}_{\tau}$ will be discovered
with 250 $fb^{-1}$ integrated luminosity and $\lambda_{311}^{\prime}=\lambda_{312}=\lambda_{321}\geq0.0093$.
If $\widetilde{\nu}_{\tau}$ mass value is not 5 TeV, $\widetilde{\nu}_{\tau}$
will be excluded with $\lambda_{311}^{\prime}=\lambda_{312}=\lambda_{321}\geq0.0045$
for $\mathcal{L}_{int}=250\,fb^{-1}$. Coupling constants limits are
summarized in Tab. \ref{tab:Yukawa-couplings-constant} for Phase
I + II upper and lower integrated luminosity values. 

\begin{table}
\caption{Yukawa couplings constant limits at different integrated luminosities.\label{tab:Yukawa-couplings-constant} }
\begin{tabular*}{16cm}{@{\extracolsep{\fill}}cc|c|c|c|c|c|c}
\hline 
\multicolumn{2}{c|}{$M_{\widetilde{\nu}_{\tau}}$(TeV)} & \multicolumn{2}{c|}{5} & \multicolumn{2}{c|}{10} & \multicolumn{2}{c}{15}\tabularnewline
\hline 
\multicolumn{2}{c|}{$\mathcal{L}_{int}(fb^{-1})$} & 250 & 17500 & 250 & 17500 & 250 & 17500\tabularnewline
\hline 
\multirow{3}{*}{$\lambda'_{311}=\lambda_{312}=\lambda_{321}$} & $2\sigma$ & $\geq$0.0045 & $\geq$0.0013 & $\geq$0.014 & $\geq$0.0024 & $\geq$0.038 & $\geq$0.005\tabularnewline
\cline{2-8} 
 & $3\sigma$ & $\geq$0.0061 & $\geq$0.0015 & $\geq$0.021 & $\geq$0.0032 & $\geq$0.057 & $\geq$0.007\tabularnewline
\cline{2-8} 
 & $5\sigma$ & $\geq$0.0093  & $\geq$0.0020 & $\geq$0.035 & $\geq$0.0048 & $\geq$0.096 & $\geq$0.012\tabularnewline
\hline 
\end{tabular*}
\end{table}

\section{Summary and Conclusions\label{sec:Conclusion}}

A search for RPV $\widetilde{\nu}_{\tau}$ neutrino decaying into
$e\mu$ final state has been carried out at the FCC-pp with its planned
very high luminosity two phases over 25 years operation time. Let
us remind that in numerical calculation $\lambda_{311}^{\prime}=0.11$
and $\lambda_{312}=\lambda_{321}=0.07$ were chosen and distribution
cuts were applied. As it can be seen at end of the first year of the
Phase I ($\mathcal{L}_{int}=250\,fb^{-1}$), $\widetilde{\nu}_{\tau}$
will be discovered ($5\sigma$), observed ($3\sigma$) and excluded
up to 15.8, 17.8 and 21.2 TeV mass values, respectively. On the other
hand, at the end of the whole 25 years operation time, the FCC-pp
will reach the highest luminosity value, $\mathcal{L}_{int}=17500\,fb^{-1}$,
$\widetilde{\nu}_{\tau}$ mass limits will be extent as 28.8 TeV for
$5\sigma$, 32.0 TeV for $3\sigma$ and 34.5 TeV for $2\sigma$ confidence
levels. 

Yukawa coupling constants, $\lambda_{311}^{\prime}$, $\lambda_{312}$
and $\lambda_{321}$, could have values instead of 0.11 or 0.07. According
to Yukawa coupling constant search over whole luminosity spectra of
the FCC (see Fig. \ref{fig:ucu-yanyana.} and Tab. \ref{tab:Yukawa-couplings-constant}),
$\widetilde{\nu}_{\tau}$ discovery at 5 TeV will be occur if $\lambda_{311}^{\prime}$,
$\lambda_{312}$ and $\lambda_{321}$ coupling constants $\geq$ 0.0020
at 17500$fb^{-1}$ integrated luminosity. On the other hand, exclusion
limits of the coupling constants are $\lambda_{311}^{\prime}=\lambda_{312}=\lambda_{321}\geq0.0013$
at the same mass value and integrated luminosity. 

It is obviously seen from the results of this work, if the $\widetilde{\nu}_{\tau}$
is exist as expected by BSM models, the next generation energy-frontier
machine, FCC-pp has a great potential to observe this SUSY particle
via examining RPV interactions. Additionally, FCC-pp gives another
opportunity to examine $\widetilde{\nu}_{\tau}$ via RPV interactions
at very low Yukawa coupling constants. 

\begin{acknowledgments}
This work was supported by the Scientific Research Project Coordination Unit of Istanbul University, Project number: BEK-2017-25593. Authors are grateful to the Usak University, Energy, Environment and Sustainability Applications and Research Center for their support. 
\end{acknowledgments}
\bibliographystyle{apsrev4-1}
\bibliography{ScalarTau}

\end{document}